\RequirePackage{fix-cm}
\documentclass[smallextended]{svjour3}       % onecolumn (second format)
\smartqed  % flush right qed marks, e.g. at end of proof
\usepackage{graphicx}
\usepackage{amssymb}
\usepackage{txfonts}
\usepackage{xcolor}
\begin{document}

\title{Study of the non-linear dynamics of micro-resonators based on a Sn-whisker in vacuum and at mK temperatures}
	


\titlerunning{The non-lin. dynamics of Sn-whisker $\mu$-resonators in vacuum \& at mK temp.}        % if too long for running head

\author{Marcel \v Clove\v cko \and
		Peter~Skyba \and \\
		Franti\v{s}ek Vavrek}


\institute{Marcel \v Clove\v cko \and Peter~Skyba \and Franti\v{s}ek Vavrek\at
	Centre of Low Temperature Physics, Institute of Experimental Physics, SAS
	and P. J. \v Saf\'arik University Ko\v sice, Watsonova 47, 04001 Ko\v sice, Slovakia\\
	Tel.: +421-55-792-2201\\
	Fax: +421-55-633-62-92\\
	\email{clovecko@saske.sk}}

\date{Received: date / Accepted: date}

\maketitle

\begin{abstract}
The dynamics of micro-resonators (or any mechanical resonators) can be studied by two complementary methods allowing the measurements in two dif\mbox{}ferent domains: (i) in the frequency domain - by the frequency sweeps using cw-excitation, and (ii) in the time domain - by the pulse techniques, when the free-decay oscillations are investigated. To broaden the knowledge about the intrinsic mechanical properties of micro-resonators based on a Sn-whisker we used both methods. We show that the dynamics of the Sn-whisker can be described by a phenomenological theory of the Duf\mbox{}f\mbox{}ing oscillator. Furthermore, we present the results of theoretical analysis based on the Duf\mbox{}f\mbox{}ing's model provided in the time and frequency domains, and we show that these results are qualitatively the same with those measured experimentally.

\end{abstract}

\section{Introduction}
\label{Intro}

Following the signif\mbox{}icant technological advancements in a semiconductor device fabrication, as a direct consequence of a very popular trend to downscale integrated electronic circuits, the physical size of mechanical resonators was successfully reduced to micro- and nanometer range \cite{NEMS-Ekinci}. The motion of such resonators is usually measured by a coupled electronic circuit, thus composing a micro-electromechanical (MEMS) or nano-electromechanical systems (NEMS). Due to their high sensitivity and simple implementation, these devices have found many commercial applications (e.g. mass sensing \cite{NEMS-mass}, frequency standards \cite{NEMS-Clock}, analogue frequency f\mbox{}ilters \cite{NEMS-Filt}).

When the resonator's dimensions are reduced, the energy related to its surface states grows and becomes comparable with the energy stored in its inner volume. Simultaneously, the resonator's surface breaks several symmetries due to which non-linear ef\mbox{}fects are gained and emphasised. These ef\mbox{}fects change the resonator's dynamics and af\mbox{}fect the intrinsic processes of the energy dissipation and exchange between the resonator and surrounding thermal bath, forming new mechanisms and channels of the energy transfer \cite{NEMS-Deco1,NEMS-Deco2}. To investigate these processes, we decided to construct a novel type of micro-resonator based on readily available Sn-whiskers \cite{Whisker1}. These metallic f\mbox{}ibres with diameter up to 1-2 $\mu$m and $\sim$ 1 mm in length grows usually on the stressed surfaces of tin alloys \cite{Whisker2,Whisker3,Whisker4}. Among the principal advantages of Sn-whisker belong that it is a single crystal, characterised by a relatively smooth surface, [0 0 1] as a preferred growth direction \cite{WhiskerGrowth}, high tensile strength (720 - 880 MPa vs 220 MPa in a bulk tin) and Young's modulus close to the bulk value \cite{WhiskerMech}. Moreover, it becomes superconducting at temperature $T_\mathrm{C} = 3.72$~K. As our measurements were performed in vacuum, at temperatures $\sim 10 $~mK and magnetic f\mbox{}ield $B = 20$~mT, one can expect that our resonator will have a reasonably high Q-factor.  
 
\section{Experimental details}
\label{ExpDetail}

\begin{figure}
	\begin{center}
		\begin{tabular}{c c}
			\includegraphics[width = 0.45\textwidth]{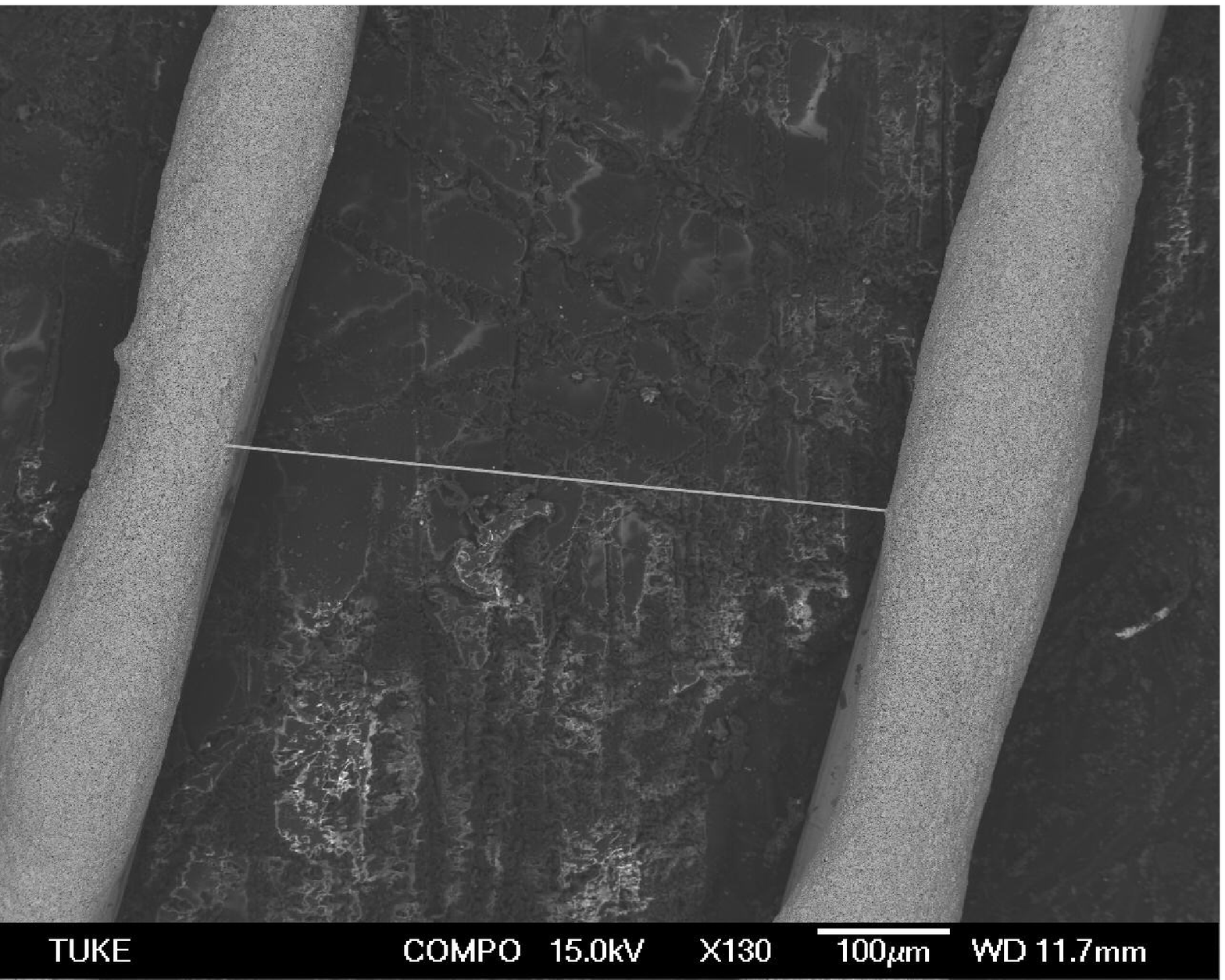} 
			&
			\includegraphics[width = 0.45\textwidth]{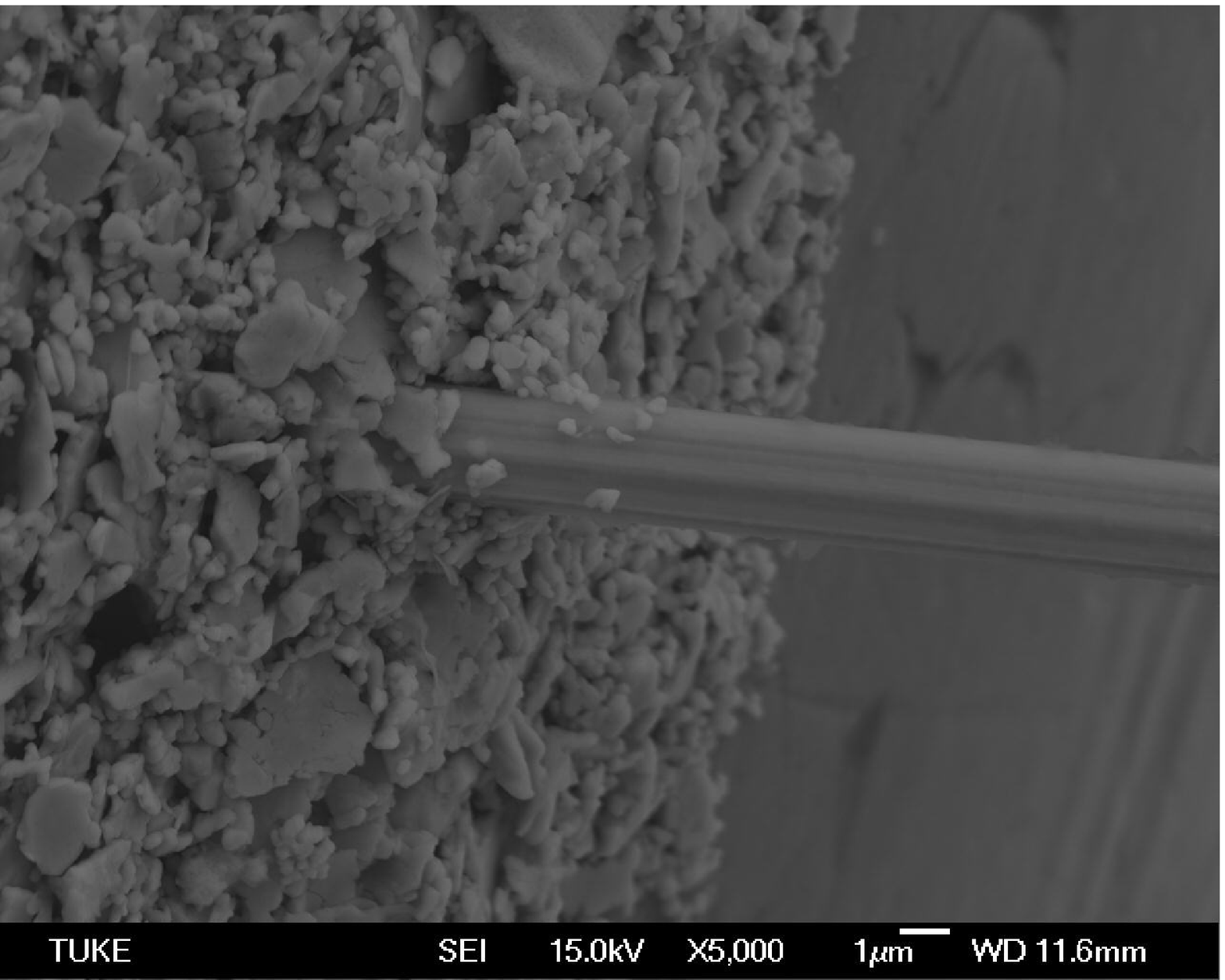}
			
		\end{tabular}
		
		%\hspace{-40pt}
		%\vspace{-8pt}
		
		\caption{SEM pictures of the micro-resonator based on Sn-whisker. A conductive silver epoxy is used to mount the whisker to two copper wires of 100 $\mu$m diameter (left). The end of Sn whisker is fully embedded in the epoxy, thus ensuring a good conductive connection (right).} 
		\label{fig1}
	\end{center}
\end{figure}

In order to construct the micro-resonator based on Sn-whisker, a relatively simple whisker holder needs to be manufactured \cite{WhiskerPaper}. Firstly, two parallel copper wires of 100 $\mu$m in diameter are glued $\sim$ 0.5 mm apart on a graph paper impregnated by the Stycast 1266 epoxy resin. The same substance is then used to attach the whole structure to a small piece of thin copper sheet with a surface insulated by a cigarette paper to ensure a good thermal and electrically non-conductive contact between both parts. The respective pairs of the copper wires are twisted and in order to protect them, a~sleeve made of nylon f\mbox{}ishing line is applied. Finally, the Sn-whisker can be positioned carefully on the f\mbox{}inished whisker holder and secured on its place by a~conductive silver epoxy resin (f\mbox{}ig.~\ref{fig1}). The whisker holder has a convenient mounting hole, so it can be mounted easily on the silver experimental plate. Placed in the middle of strong superconducting magnet and thermally connected to the mixing chamber of the commercially available cryogen free dilution refrigerator, this experimental set-up allows us to perform measurements in the magnetic f\mbox{}ields up to 8 T and at temperatures down to 10 mK. 

\begin{figure}
	\begin{center}
		\begin{tabular}{c c}
			\includegraphics[width = 0.450\textwidth]{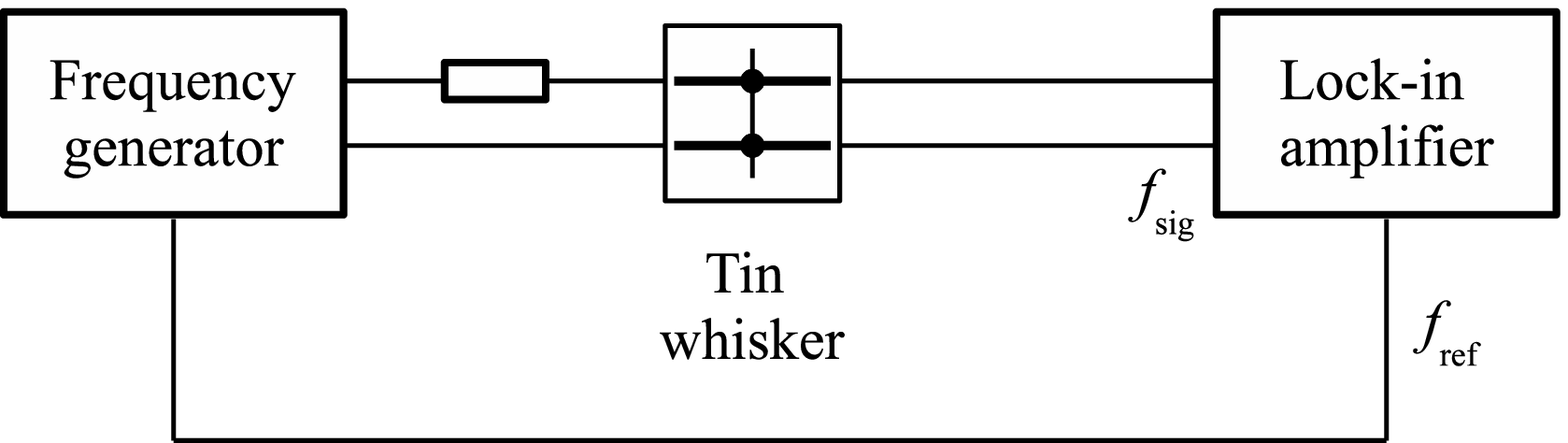}
			&
			\includegraphics[width = 0.479\textwidth]{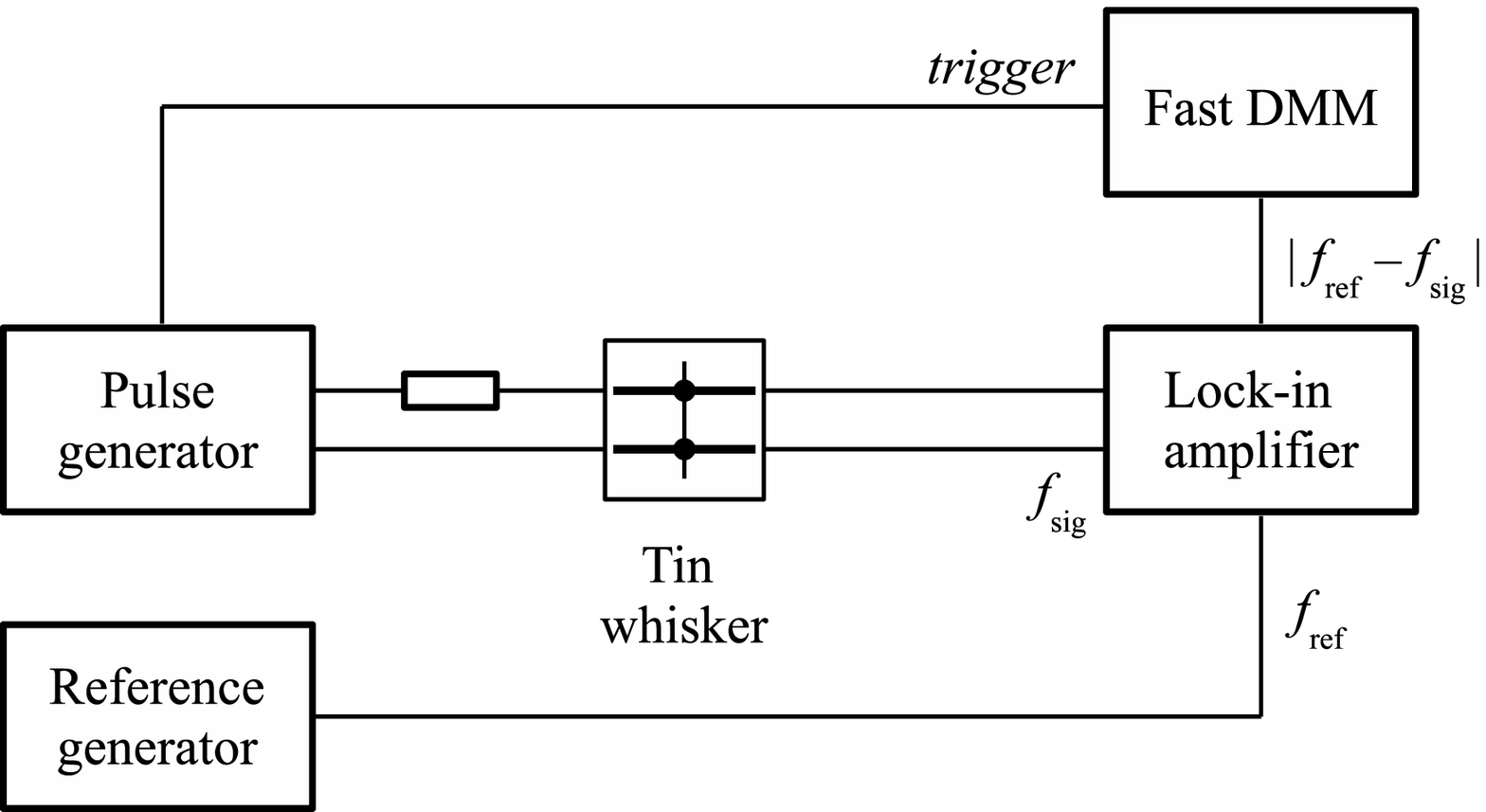}
			% text width ratio Pulse to Sweep is 1.064285714
		\end{tabular}
		
		%\hspace{-40pt}
		%\vspace{-8pt}
		
		\caption{The measurement scheme used for the traditional technique of frequency sweeps with continuous voltage excitation (left) and for the pulse-demodulation technique (right). } \label{fig2}
	\end{center}
\end{figure}

The physical properties and dynamics of micro- and nano-resonators are usually studied in the frequency domain by the traditional technique of frequency sweeps with continuous voltage/current excitation (f\mbox{}ig. \ref{fig2} \textit{left}). A precise function generator serves as a current source and the response of Sn-whisker based micro-resonator in the form of induced voltage $U_\mathrm{i} \sim B l v$ is measured by a phase-sensitive (Lock-in) amplif\mbox{}ier. Its reference signal is provided by the same function generator, so the both response components can be determined. As we are dealing with the forced oscillations driven by excitation at given frequency, the dynamics of micro-resonator based on Sn-whisker is studied in the frequency domain.

There is a complementary measurement method represented by a pulse technique when the free damped oscillations are examined in the time domain.
If the corresponding FFT spectrum is calculated, the dynamics of micro- and nano-resonators can be transformed to the respective frequency domain and results can be compared to the frequency sweeps measurements. We have utilised a novel type of the pulse-demodulation technique (f\mbox{}ig. \ref{fig2} \textit{right}) implemented successfully for the quartz tuning forks \cite{PulseTech}. In contrast to the frequency sweeps technique there is an additional precise function generator providing the reference signal at frequency $f_\mathrm{ref}$ for the Lock-in amplif\mbox{}ier. Moreover, a fast DMM is connected to the output of Lock-in amplif\mbox{}ier and used for the data acquisition and storage. At f\mbox{}irst, the excitation pulse is applied at frequency $f_\mathrm{exc}$ for N periods. At the end of excitation pulse the fast DMM is triggered and the signal of free damped oscillations (originally at $f_\mathrm{sig}$) is being measured at dif\mbox{}ferential frequency $|f_\mathrm{ref}-f_\mathrm{sig}|$. 

\section{Experimental results and discussion}
\label{ExpRes}

\begin{figure}
	\begin{center}
		\begin{tabular}{c c}
			\includegraphics[width = 0.46\textwidth]{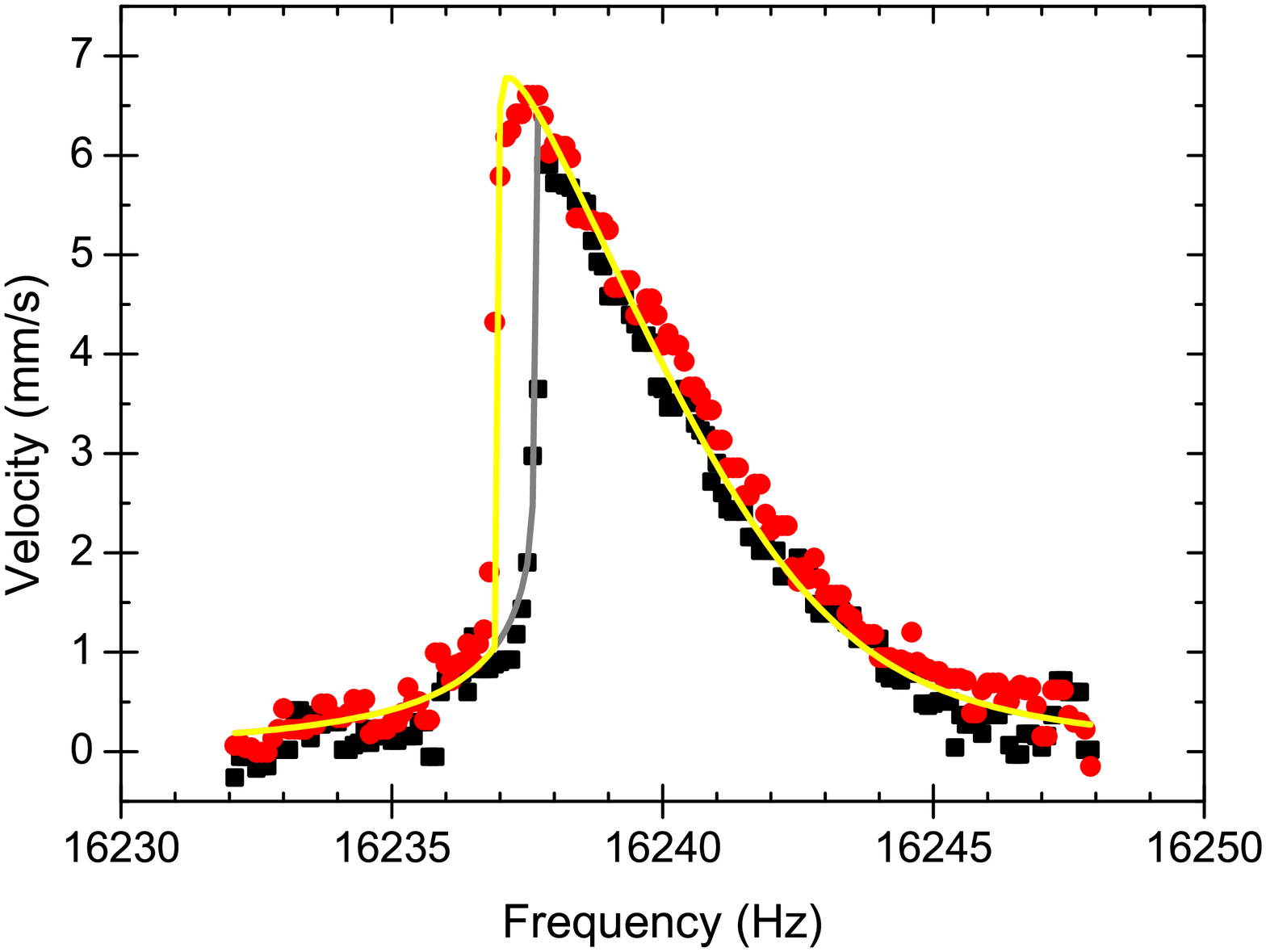} 
			&
			\includegraphics[width = 0.465\textwidth]{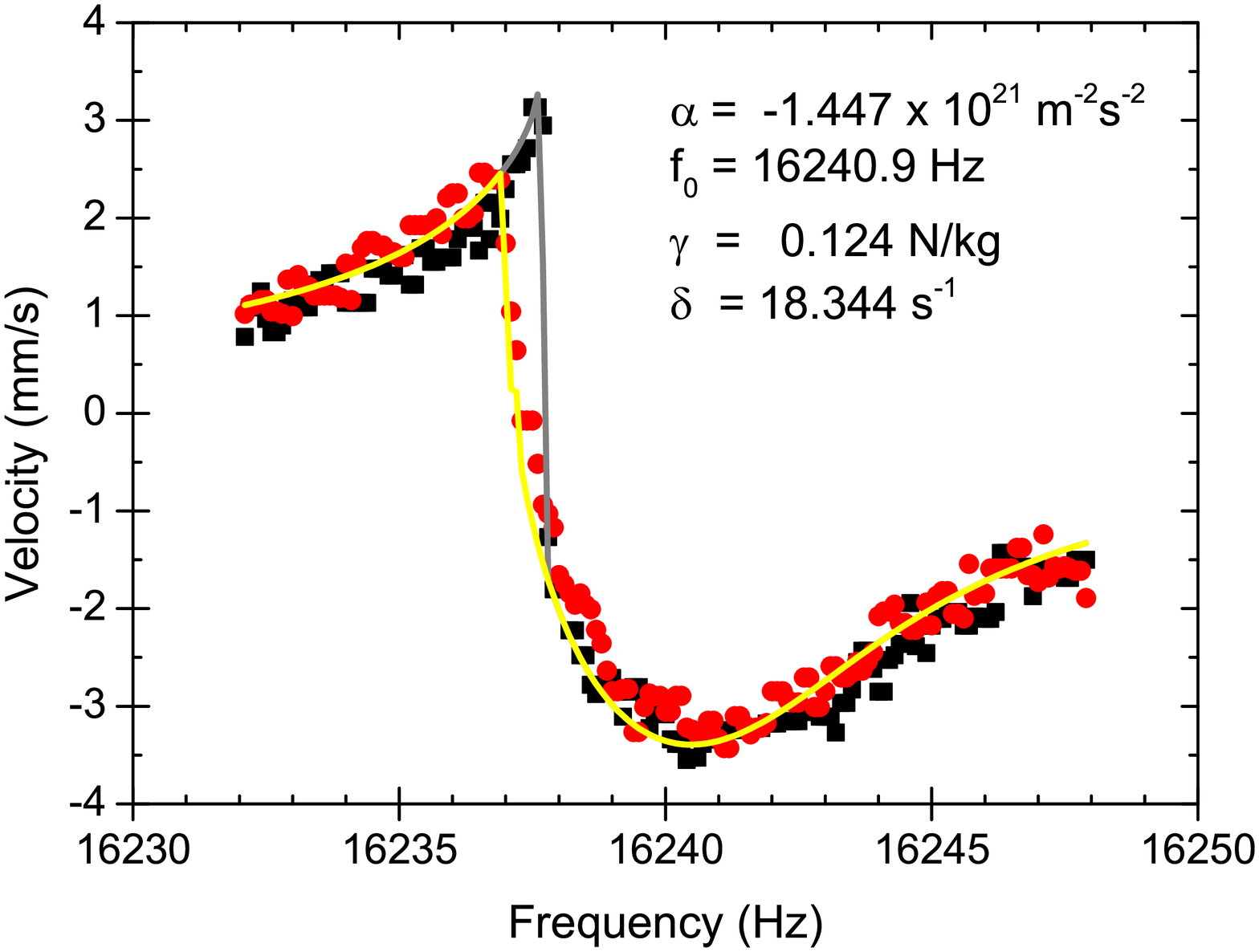}
		\end{tabular}
		
		%\hspace{-40pt}
		%\vspace{-8pt}
		
		\caption{The typical dependencies of in-phase (left) and quadrature (right) velocity components measured during up ($\blacksquare$) and down (\textcolor{red}{$\medbullet$}) frequency sweeps in the case of higher voltage excitations, a clear sign of hysteresis can be seen. The solid lines represents the f\mbox{}its based on the model of Duf\mbox{}f\mbox{}ing oscillator (see text for details).} \label{fig3}
	\end{center}
\end{figure}

%\subsection{The frequency sweeps and the frequency domain}
As it was shown during the previous frequency sweeps measurements \cite{WhiskerPaper}, the response of Sn-whisker based micro-resonator becomes non-linear even at the moderate current excitations. The resonant curve is tilted towards the lower frequencies and as the current excitation increases this tendency becomes more prominent. Our new experimental results conf\mbox{}irm that when a critical excitation is exceeded, the direction of frequency sweep starts to play an important role and as a result, the ef\mbox{}fect of hysteresis is observed (f\mbox{}ig. \ref{fig3}). Clearly, there is a frequency region where the resonator can oscillate either with high or low velocity amplitude depending on the sweep's history. 

\begin{figure}
 	\begin{center}
 		\begin{tabular}{c c}
 			\includegraphics[width = 0.455\textwidth]{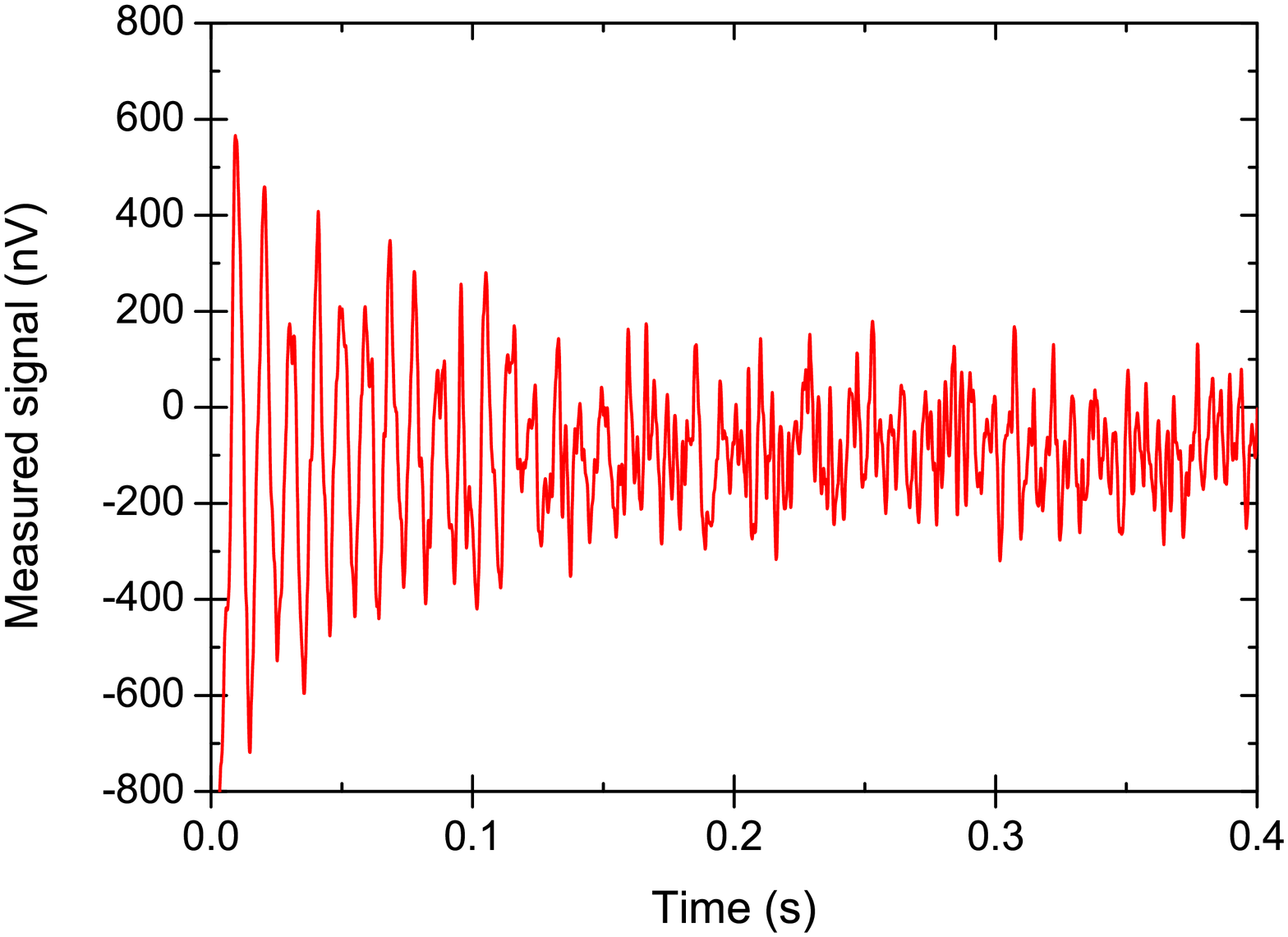}
 			&
 			\includegraphics[width = 0.468\textwidth]{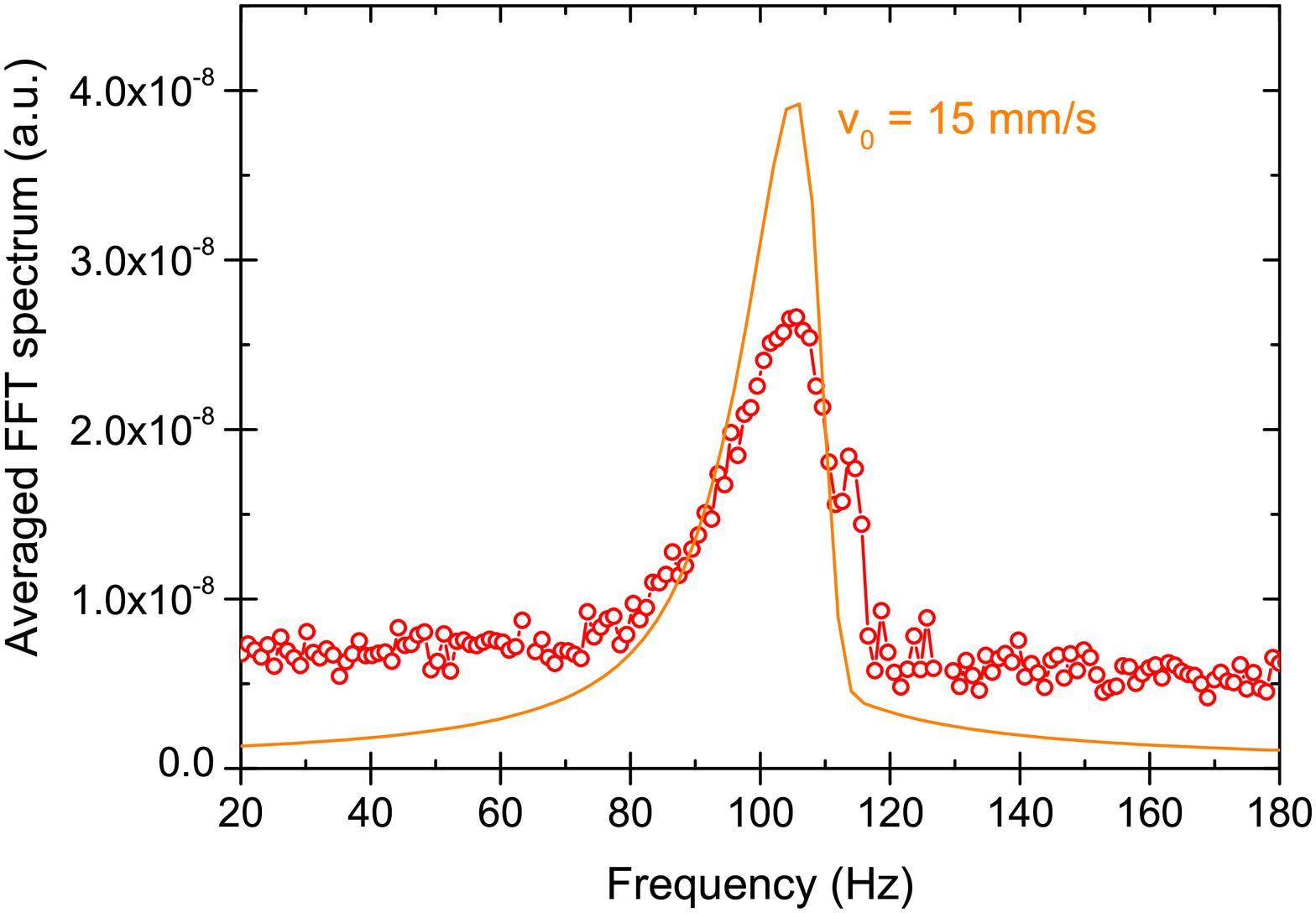}
 			% text width ratio FFT to Pulse is 1.028888889
 		\end{tabular}
 		
 		%\hspace{-40pt}
 		%\vspace{-8pt}
 		
 		\caption{A signal of free damped oscillations measured for the following excitation pulse parameters $f_\mathrm{exc} = 16\mbox{ }240$ Hz, $I_\mathrm{exc}$ and $N = 16\mbox{ }000$ (\textit{left}) and corresponding FFT spectrum averaged from 20 pulse measurements (\textit{right}). The solid line represents the FFT spectrum calculated using the model of Duf\mbox{}f\mbox{}ing oscillator with the parameters $\alpha$, $\omega_0$ and $\delta$ determined from the frequency sweep measurements.} \label{fig4}
 	\end{center}
\end{figure}

%\subsection{The time domain studied by the pulse-demodulation technique}
To deepen our understanding about the intrinsic mechanical properties of micro-resonator based on Sn-whisker, we studied its response in the time domain by the means of pulse-demodulation technique. The frequency of excitation pulse $f_\mathrm{exc} = 16\mbox{ }240$~Hz was chosen to be near the resonant frequency $f_0$. There are two parameters of excitation pulse which can be adjusted in order to obtain a reasonable signal-to-noise ratio: the number of periods $N$ and the pulse amplitude $I_\mathrm{exc}$. It is worth to note that due to a low signal-to-noise ratio the amplitude of excitation pulse was signif\mbox{}icantly larger ($\sim 20 \:\times$) than during the frequency sweeps measurements. Similarly, the number of periods was set to $N = 16\mbox{ }000$. The resulting signal of free damped oscillations is shown in a f\mbox{}ig. \ref{fig4} \textit{left}. To improve the corresponding FFT spectrum, the pulse measurement for given parameters $f_\mathrm{exc}$, $I_\mathrm{exc}$ and $N$ was repeated 20~times, individual FFT spectra were determined and, f\mbox{}inally, the average FFT spectrum was calculated (f\mbox{}ig. \ref{fig4} \textit{right}). As the signal of free damped oscillations is being measured at dif\mbox{}ferential frequency $|f_\mathrm{ref}-f_\mathrm{sig}|$ the whole FFT spectrum is transposed to the lower frequency range. Moreover, the FFT spectrum is not symmetrical and higher frequencies are more pronounced.

%\subsection{A phenomenological theory of a Duffing oscillator}
The response of Sn-whisker based micro-resonator during the frequency sweeps measurements resembles the behaviour of Duf\mbox{}f\mbox{}ing oscillator, especially at higher current excitations. In general, the motion of forced Duf\mbox{}f\mbox{}ing oscillator can be described by a following ordinary dif\mbox{}ferential equation
\begin{equation}
\label{DuffingOsc}
\ddot{x} + \delta \dot{x} + \omega_0^2 \:\! x + \alpha x^3 = \gamma \cos (\omega t) \,,
\label{eqDuffing}
\end{equation}
where $\delta$ represents damping, $\omega_0$ is the resonant angular frequency, $\alpha$ determines the resulting non-linearity of the restoring force and $\gamma$ is the amplitude of a harmonic driving force normalized per unit mass and acting at the angular frequency $\omega$. This equation is not analytically solvable; however, there are many mathematical methods available which allow to f\mbox{}ind an approximate steady-state solution. Using the homotopy analysis \cite{HomotopyAnal} or harmonic balance method \cite{HarmBalance}, it is possible to derive the expressions for displacement amplitude $r$ and phase $\phi$ 
\begin{eqnarray}
 r^2 & = & \phantom{-} \frac{\gamma^2}{\left(\omega^2 - \omega^2_0 - \frac{3}{4}\:\! \alpha \:\! r^2\right)^2 + \omega^2\delta^2} \,, \label{eqDisAmpl} \\
 \sin \phi & = & -\frac{\omega \:\! \delta r}{\gamma} \,. \label{eqDisPhase}
\end{eqnarray}
As a side note, there is a special case $\alpha = 0$, when the frequency response of Duf\mbox{}f\mbox{}ing oscillator has the same shape of Lorentzian function as for the linear harmonic oscillator. The same is true for the case of small driving forces $\gamma \ll 1$ when $\frac{3}{4}\:\! \alpha \:\! r^2 \to 0$, i.e. in the linear regime, then the quality factor Q of our micro-resonator can be determined as well. The result $Q = \frac{\omega_0}{\delta} \sim 5500$ is in the excellent agreement with our previous measurements \cite{WhiskerPaper}.
  
By analysing the equation (\ref{eqDisAmpl}) for the displacement amplitude $r$, it is possible to determine the angular frequency $\omega_\mathrm{max}$ when the response of Duf\mbox{}f\mbox{}ing oscillator is maximal
 \begin{equation}
 \omega_\mathrm{max}^2 = \omega^2_0 + \frac{3}{4}\:\! \alpha \:\! r^2 \,.
 \end{equation}
As one can see, the response of Duf\mbox{}f\mbox{}ing oscillator at the higher amplitudes of displacement $r$ is no longer linear. The resonance curve is not symmetrical any more and depending on the sign of $\alpha$, there is an ef\mbox{}fect of softening (for $\alpha < 0$) or hardening (for $\alpha > 0$) observed. Moreover, at even larger displacement amplitudes, the response of oscillator will depend on the direction of frequency sweep itself and is characterised by a hysteresis behaviour with distinctive abrupt changes in the amplitude of oscillations.
 
In principle, the expression (\ref{eqDisAmpl}) is a cubic equation for $r^2(\omega)$ and can be solved numerically for the given parameters $\alpha$, $\omega_0$, $\gamma$ and $\delta$. Once a computational algorithm for the calculation of $r\:\!(\omega)$ is available, the relation (\ref{eqDisPhase}) can be used to determine $\phi(\omega)$. Moreover, by implementing the well-known Levenberg-Marquardt method \cite{Levenberg,Marquardt,Leven-Marq}, it is possible to construct a software (numerical) f\mbox{}itting function for the experimentally measured $v(\omega)$ dependencies. In our case, the coef\mbox{}f\mbox{}icients $\alpha$, $\omega_0$, $\gamma$ and $\delta$ serve as the initial guess parameters for our f\mbox{}itting function. In addition, the experimental data from the up and down frequency sweeps are being processed at the same time and, f\mbox{}inally, the best f\mbox{}it parameters are determined. The computed results (represented by the solid lines in f\mbox{}ig. \ref{fig3}) conf\mbox{}irm a def\mbox{}inite agreement between the measured experimental data and calculated f\mbox{}it. 

Now, we apply the model of Duf\mbox{}f\mbox{}ing oscillator for the theoretical description of the free damped oscillations of Sn-whisker based micro-resonator, with parameters $\alpha$, $\omega_0$ and $\delta$ determined from the frequency sweeps measurements. In this case, the right-hand side of equation (\ref{eqDuffing}) is equal to zero (no driving force is applied). In order to simplify the necessary numerical calculations, the fourth-order Runge-Kutta method was chosen with the initial conditions selected as $x_{\mathrm{t} \, = \, 0} = 0$ and $v_{\mathrm{t} \, = \, 0} = v_0$. The resulting signal was calculated up to 0.5 s time interval and the corresponding FFT spectrum was determined and then transposed to the lower frequency range by subtracting the reference frequency $f_\mathrm{ref}$. For initial velocity $v_0 = 15$~mm/s, there is a~reasonable qualitative agreement between measured and simulated FFT spectra (see f\mbox{}ig. \ref{fig4} \textit{right}, the solid line represents the simulated FFT spectrum). Knowing the initial induced voltage amplitude $U_\mathrm{i} \sim 600$~nV (see f\mbox{}ig. \ref{fig4} \textit{left}), it is possible to estimate $v_0$ as well. For the whisker of $\sim 1$~mm length placed in the magnetic f\mbox{}ield of 20 mT, the resulting order of magnitude estimate for the initial velocity is $v_0 \sim 30$~mm/s. This agrees quite well with the simulated value of 15 mm/s. %To figure out, why there is only the qualitative agreement, more experiments need to be done in future.

\section{Conclusions}
\label{Conclusions}

To conclude, we studied the dynamics of micro-resonator based on Sn-whisker by measuring its response in both (frequency and time) domains. Besides the traditional measurement method - the frequency sweeps using continuous excitation, the modif\mbox{}ied pulse-demodulation technique was proposed and implemented. Moreover, the phenomenological theory of Duf\mbox{}f\mbox{}ing oscillator applied to the micro-resonator's response in the frequency domain resulted in the excellent qualitative agreement between the experimental data and corresponding f\mbox{}its. Using the known values of f\mbox{}itting parameters $\alpha$, $\omega_0$ and $\delta$, obtained from the implemented Duf\mbox{}f\mbox{}ing's theory, the free damped oscillations were calculated and the corresponding FFT spectra were compared to the averaged FFT spectra determined from the resonator's response in the time domain. As it was shown, there is a qualitative agreement between them. Thus, we have proved a one-to-one qualitative correspondence between measurements in time and frequency domains.

%Your text comes here. Separate text sections with
%\section{Section title}
%\label{sec:1}
%Text with citations \cite{RefB} and %\cite{RefJ}.
%\subsection{Subsection title}
%\label{sec:2}
%as required. Don't forget to give each %section
%and subsection a unique label (see Sect.~\ref{sec:1}).
%\paragraph{Paragraph headings} Use paragraph headings as needed.
%\begin{equation}
%\end{equation}

%\begin{figure}
%  \includegraphics{example.eps}
%\end{figure}
%\begin{figure*}
%  \includegraphics[width=0.75\textwidth]{example.eps}
%\end{figure*}

%\begin{table}
%\begin{tabular}{lll}
%\hline\noalign{\smallskip}
%\end{tabular}
%\end{table}

\begin{acknowledgements}
We would like to thankfully acknowledge the support by grants APVV 14-0605, VEGA 2/0157/15 and European Microkelvin Platform.
The f\mbox{}inancial support provided by the U. S. Steel Ko\v sice s.r.o. is also gratefully recognised and highly appreciated.

\end{acknowledgements}


\begin{thebibliography}{}

\bibitem{NEMS-Ekinci} K. L. Ekinci, M. L. Roukes, \textit{Nanoelectromechanical systems}, Rev. Sci. Instrum. 76, 061101 (2005)

\bibitem{NEMS-mass} J. Chaste, A. Eichle, J. Moser, G. Ceballos, R. Rurali, A. Bachtold, \textit{A nanomechanical mass sensor with yoctogram resolution}, Nature Nanotechnol. 7, 301 (2012)

\bibitem{NEMS-Clock} C. Lam, \textit{A review of the recent development of MEMS and crystal oscillators and their impacts on the frequency control products industry}, Proc. IEEE Ultrasonics Symp., pp. 694 – 704 (2008)

\bibitem{NEMS-Filt} S.-S. Li, Y.-W. Lin, Z. Ren, and C.-C. Nguyen, \textit{An MSI Micromechanical Dif\mbox{}ferential Disk-Array Filter}, Solid-State Sensors, Actuators and Microsystems Conference 2007, pp. 307 – 311 (2007)

\bibitem{NEMS-Deco1} L. G. Remus, M. P. Blencowe, Y. Tanaka, \textit{Damping and decoherence of a nanomechanical resonator due to a few two-level systems}, Phys. Rev. B 80, 174103 (2009)

\bibitem{NEMS-Deco2} O. Maillet, F. Vavrek, A. D. Fef\mbox{}ferman, O. Bourgeois, E. Collin, \textit{Classical decoherence in a nanomechanical resonator}, New J. Phys. 18, iss. 7, 073022 (2016)


%\bibitem{C-nanotubes1} S. Sapmaz, Y. M. Blanter, L. Gurevich, H. S. J. van der Zant, \textit{Carbon nanotubes as nanoelectromechanical systems}, Phys. Rev. B 67, pp. 235414 (2003)

%\bibitem{C-nanotubes2} A. K. H\" uttel, G. A. Steele, B. Witkamp, M. Poot, L. P. Kouwenhoven, H. S. J. van der Zant, \textit{Carbon Nanotubes as Ultrahigh Quality Factor Mechanical Resonators}, Nano Lett. 9, pp. 2547 (2009)

\bibitem{Whisker1} K. G. Compton, A. Mendizza, S. M. Arnold, \textit{Filamentary growths on metal surfaces—“whiskers”}, Corrosion
7, 327 (1951)

\bibitem{Whisker2} R. M. Fisher, L. S. Darken, K. G. Carroll, \textit{Accelerated growth of tin whiskers}, Acta Metal. 2, 368 (1954)

\bibitem{Whisker3} G. T. Gaylon, \textit{A History of a Tin Whisker Theory: 1946 to 2004}, iNEMI. Freely available on http://thor.
inemi.org/webdownload/newsroom/Presentations/SMTAI-04\_tin\_whiskers.pdf

\bibitem{Whisker4} Y. Sun, E. N. Hof\mbox{}fman, P.-S. Lam, X. Li, \textit{Evaluation of local strain evolution from metallic whisker formation}, Scr. Mater. 65, 388 (2011)

\bibitem{WhiskerGrowth} W. J. Choi, T. Y. Lee, K. N. Tu, N. Tamura, R. S. Celestre, A. A. MacDowell, Y. Y. Bong, Luu Nguyen, \textit{Tin
whiskers studied by synchrotron radiation scanning X-ray micro-dif\mbox{}fraction}, Acta Materialia 51, 6253 (2003)

\bibitem{WhiskerMech} S. S. Singh, R. Sarkar, H.-X. Xie, C. Mayer, J. Rajagopalan, N. Chawla, \textit{Tensile Behavior of Single-Crystal Tin
Whiskers}, J. Electron. Mater. 43 (4), 978 (2014)

\bibitem{WhiskerPaper} M. \v Clove\v cko, E. Ga\v zo, S. Longauer, E. M\' udra, P. Skyba, F. Vavrek, M. Vojtko, \textit{Vacuum Measurements of a Novel Micro-resonator Based on Tin Whiskers Performed at mK Temperatures}, J. Low Temp. Phys. 175, 449 (2014)

\bibitem{PulseTech} M. \v Clove\v cko, M. Grajcar, M. Kupka, P. Neilinger, M. Reh\'ak, P. Skyba, F. Vavrek, \textit{High Q value Quartz Tuning Fork in Vacuum as a Potential Thermometer in Millikelvin Temperature Range}, J. Low Temp. Phys. 187, 573 (2017)

\bibitem{HomotopyAnal} F. Tajaddodianfar, M. R. H. Yazdi, H. N. Pishkenari, \textit{Nonlinear dynamics of MEMS/NEMS resonators: analytical solution by the homotopy analysis method}, Microsystem Technologies. 23, 1913 (2017) https://doi.org/10.1007/s00542-016-2947-7

\bibitem{HarmBalance} D. W. Jordan, P. Smith, \textit{Nonlinear ordinary dif\mbox{}ferential equations – An introduction for scientists and engineers (4th ed.)}, pp. 223–233, Oxford University Press, (2007) ISBN 978-0-19-920824-1

\bibitem{Levenberg} K. Levenberg, \textit{A Method for the Solution of Certain Problems in Least Squares}, Quart. Appl. Math.~2, pp. 164-168 (1944)

\bibitem{Marquardt} D. Marquardt, \textit{An Algorithm for Least-Squares Estimation of Nonlinear Parameters}, SIAM J. Appl. Math. 11, pp. 431-441 (1963)

\bibitem{Leven-Marq} P. R. Gill, W. Murray, M. H. Wright, \textit{The Levenberg-Marquardt Method}, §4.7.3 in Practical Optimization. London: Academic Press, pp. 136-137 (1981)








% etc
\end{thebibliography}


\end{document}